\documentclass{article}
\usepackage{spconf,amsmath,graphicx,hyperref}

\title{Empowering Economic Simulation through \\Situation-aware LLM-driven Generative System}
\name{Zhimei Chen$^{1}$ \quad Mu Chen$^{2*}$\thanks{Corresponding author: mu.chen@alumni.uts.edu.au}}
\address{$^{1}$ School of Economics, Southwest Minzu University \\
         $^{2}$ ReLER, AAII, University of Technology Sydney}

\usepackage{booktabs}
\usepackage{multirow}
\usepackage[table]{xcolor}
\usepackage{colortbl}
\usepackage{graphicx}
\usepackage{caption}
\usepackage{subcaption}
\usepackage{dashrule}
\usepackage{enumitem}
\usepackage{hyperref}
\usepackage{pifont}
\definecolor{mygray}{gray}{.93}

\usepackage{xspace}

\begin{document}
\maketitle
\begin{abstract}
Traditional economic modeling typically follows a \textsc{top-down} paradigm, neglecting individual diversity and the complexity of social interactions. To better capture the complexity of societal structure, Agent-Based Modeling (ABM) employs a \textsc{bottom-up} solution by incorporating micro-level dynamics to generate macroeconomic phenomena. Reinforcement Learning further improves its decision-making ability through tailored reward signals. However, existing ABM systems struggle to generalize beyond predefined scenarios. Recognizing the potential of LLM-driven role-playing in perception and human-like decision-making, we propose \textsc{SaMAS}, which models individual agents with rich macroeconomic understanding embedded in LLMs and economic trajectories experienced in the passing simulation steps. By jointly modeling both macro-level structural patterns and micro-level dynamic behaviors, \textsc{SaMAS} achieves superior performance in volatility realism and turning point prediction.

\end{abstract}

\vspace{+12pt}
\section{Introduction}
\label{sec:intro}

   \begin{figure}[t]
      \begin{center}
     \includegraphics[width=1.\linewidth]{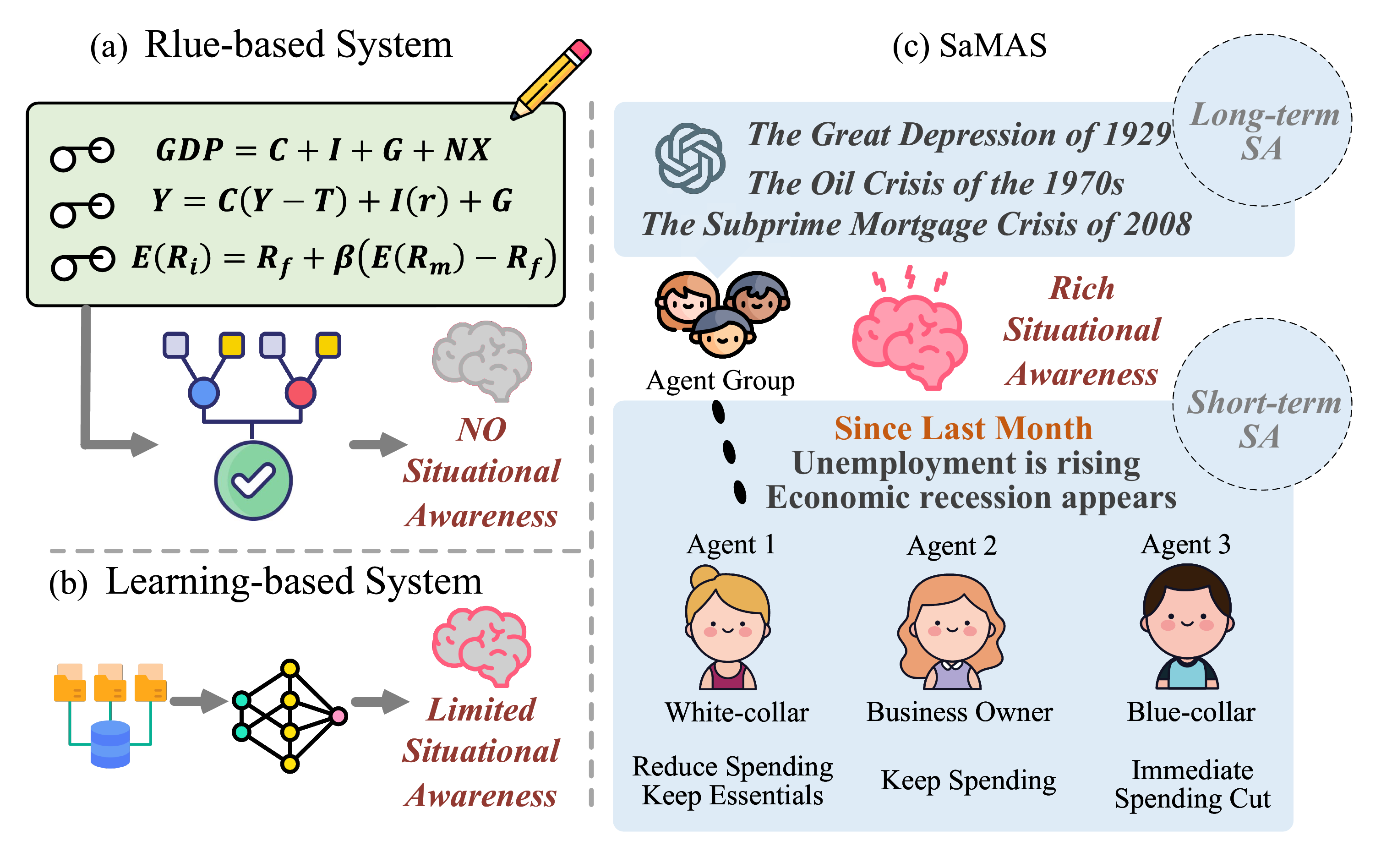}
          \end{center}
      \vspace{-8pt}
      \captionsetup{font=small}
      \caption{\small{(a) Existing economic simulation systems adopt a rule-based paradigm, exhibiting no situational awareness. (b) Learning-based simulation systems learn from historical data and acquire limited situational awareness by understanding various economic contexts.  (c)\textsc{SaMAS} explicitly leverages underlying macroeconomic knowledge stored in LLMs and stores micro situations experienced by each simulation agent, leading to rich situational awareness.
      }
      }
      \label{fig:1}
            \vspace{-0pt}

    \end{figure}

    \begin{figure*}[t]
        \begin{center}
            \includegraphics[width=1\linewidth]{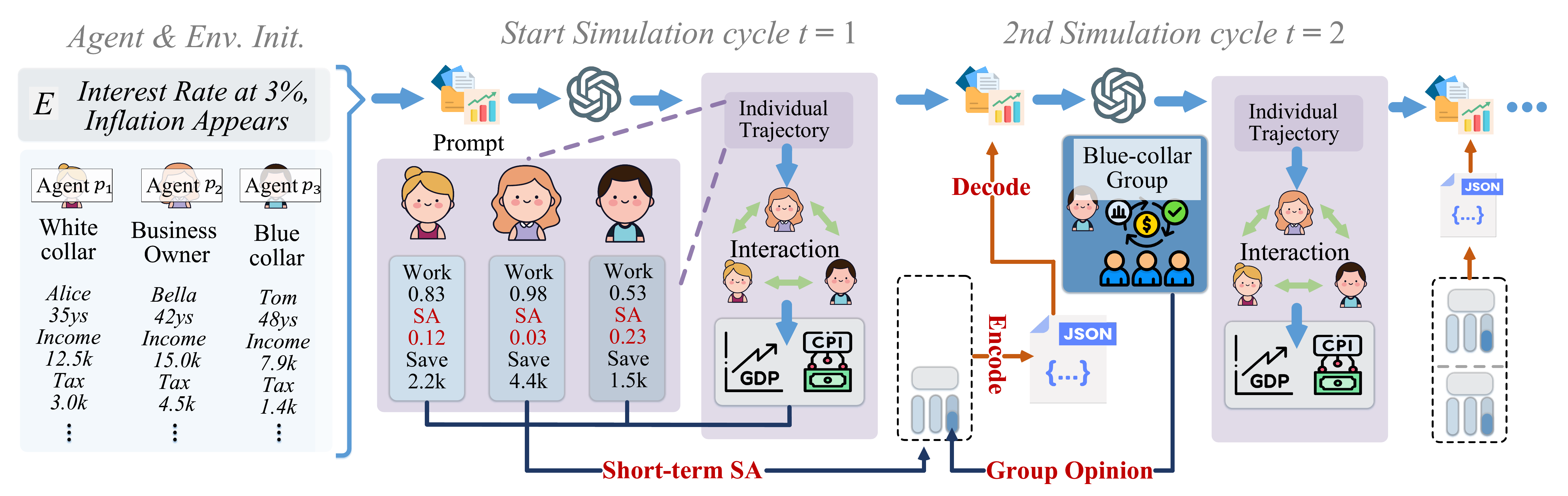}
            \end{center}
        \vspace{-12pt}
        \captionsetup{font=small}
        \caption{\small{\textsc{SaMAS} equips with long-term and short-term awareness, the former is from the commonsense awareness in LLMs' parameters, and the latter is from stored individual trajectories of agents in the system.}
        }
        \label{fig:2} 
        \vspace{-12pt}
    \end{figure*}

Classical economic modeling normally adopts a \textsc{top-down} paradigm,$_{\!}$ based$_{\!}$ on$_{\!}$ either$_{\!}$ a$_{\!}$ theory-driven~\cite{smets2007shocks},$_{\!}$ or a data-driven method~\cite{sims1980macroeconomics}.$_{\!}$ These \textsc{top-down} solutions have served as the foundation$_{\!}$ of macroeconomic analysis and policy formulation. While effective in capturing empirical correlations among macroeconomic variables, the passive introduction of the ``rational expectation assumption''~\cite{muth1961rational} neglects the diversity of individuals in society and the complexity of social interaction, leading to an unrealistic social structure. Agent-Based Modeling (ABM)~\cite{tesfatsion2002agent,farmer2009economy} 
addresses this by simulating individual agent behaviors and the interactions among them. By considering micro-level dynamics, ABM generates macroeconomic phenomena in a \textsc{bottom-up} manner. Reinforcement learning~\cite{rashid2020monotonic} further extends this direction by enabling agents to learn adaptive strategies from interaction with the environment via reward signals, demonstrating strong performance in complex social simulation settings.

Despite the impressive decision-making capabilities brou-\\ght by RL-based ABM, they adopt \textsc{tailored} reward functions to mimic diverse human behavior. As a result, they fall short in generalizing beyond pre-defined scenarios. Actually, humans rely on past experiences to guide current decisions. Specifically, the perception of economic environments shapes humans' risk preferences~\cite{schildberg2018risk}. For instance, as shown in Fig~\ref{fig:1}, awareness of historical crises such as \textit{the Great Depression of 1929}, \textit{the Oil Crisis of the 1970s}, and \textit{the Subprime Mortgage Crisis of 2008} tends to trigger panic when similar economic signals emerge. On the other hand, individuals are easily affected by recent successful financial decisions~\cite{cakici2023recency,kotomin2025debiasing}, termed \textsc{recency bias}. Furthermore, individuals gradually develop similar economic decision patterns~\cite{hill2019choice}, forming group decisions. Specifically, the economic activities of different individuals are highly correlated to their professions, as they share: \ding{182} similar income and consumption structures; \ding{183} homogeneous information propagates among them in their daily interactions~\cite{sunstein2001echo}.
For example, white-collar professionals tend to reduce unnecessary spending but maintain essential consumption during recessions to mitigate unemployment risks. In contrast, business owners keep spending since they typically hold larger savings. Blue-collar workers, however, will immediately cut back on spending due to the lack of savings. Current ABM solutions could not capture these intrinsic microeconomic patterns, which limits their ability to reproduce group-level decision dynamics.

The generative nature of Large Language Models (LLMs) has recently made them attractive as role-playing agents, as they can satisfy the core ABM requirement of equipping individual agents with perception and decision-making capabilities. The macroeconomic understanding encoded in an LLM's parameters provides a rich background context for an individual agent’s long-term awareness, such as prior understanding of financial crises. Nevertheless, humans' economic decisions not only rely on long-term historical knowledge but also on their recent experiences~\cite{wang2025unable}. Effective decision-making thus requires dynamically balancing long-term macroeconomic understanding and short-term contextual observations. With such insight, we propose a Situation-aware Multi-agent System (\textsc{SaMAS}) for realistic economic simulation. Within the system, two paradigms of economic awareness predominate: Long-term Awareness (LA) and Short-term Awareness (SA). The former builds upon the historical understanding embedded in LLMs, while the latter derives from individual-specific trajectories within the recent context window. The two paradigms provide complementary perspectives, capturing macro-level social consensus and micro-level adaptive behaviors, respectively. \textsc{SaMAS} demonstrates superior Volatility Realism and Hit Rate compared to existing methods.

\vspace{-6pt}

\section{Related Work}
\label{sec:format}
\noindent\textbf{Traditional Economic Modeling} primarily relies on statistical and equilibrium-based frameworks to analyze aggregate behaviors and macroeconomic patterns~\cite{smets2007shocks,sims1980macroeconomics}, including representative  Dynamic Stochastic General Equilibrium (DSGE)  and Vector Autoregression (VAR), serving as representative examples that are widely applied in policymaking. A key limitation of these approaches, however, is their reliance on stylized assumptions such as agent homogeneity and perfect rationality, which substantially oversimplify the complexity of real-world economies. In particular, such assumptions neglect individual heterogeneity, bounded rationality, and the social network structures that shape economic outcomes.

\noindent\textbf{Agent-Based Modeling} explains macro-level outcomes through micro-level interactions~\cite{tesfatsion2002agent,farmer2009economy}. ABM simulates the behaviors of heterogeneous agents and their localized interactions, thereby partially addressing the limitations of traditional macroeconomic modeling, which often neglects individual heterogeneity and complex interaction structures~\cite{bonabeau2002agent}.  Nevertheless, simplified ABM frameworks are deficient in situational awareness, which limits their ability to capture the subtle and context-dependent characteristics of human economic decision-making. To mitigate this, recent research has increasingly explored the integration of RL~\cite{chen2025seed} and DL~\cite{chen2023pipa,chen2024transferring,chen2024general,chen2025diffvsgg,chen2024pipa++} into ABM frameworks to enhance intelligence. In particular, RL empowers agents to iteratively learn strategies through interaction with their environment.

\noindent\textbf{Situational Awareness} is also correlated with the proposed system. The emergence of large generative models has revealed unprecedented potential for enabling agents to perceive, reason, and make decisions. Contextual awareness plays a pivotal role in these processes, as it directly informs perception and guides decision-making. For socially situated agents, memory functions not only as a repository of historical information but also as the foundation for cognitive consistency, situational understanding, and the development of long-term strategies. Existing approaches, however, largely rely on either external memory augmentation, such as retrieval-augmented generation (RAG)~\cite{lewis2020retrieval}, or reinforcement learning-based mechanisms to control memory access~\cite{xu2025mem}. While effective in short-term interaction tasks, these methods remain inadequate for macroeconomic simulation, which is inherently characterized by long-term temporal evolution. In particular, current methods fall short in modeling multi-timescale dynamics and the complex interaction structures spanning environments, individuals, and groups.

\vspace{-6pt}
\section{\textsc{SaMAS}}
\textsc{SaMAS} aims to construct a more realistic social environment by leveraging the generative and reasoning capacities of LLMs in role-playing agents. It addresses the simulation pipeline at two complementary levels: (\textbf{micro}) the decision-making processes of individual agents, and (\textbf{macro}) the emergent dynamics arising from their collective interactions.
\vspace{-6pt}

\subsection{Overview}
As shown in Fig~\ref{fig:2}, consistent with the principles of ABM, the overall economic environment is generated through the interactions of individual role-playing agents, and the overall simulation performance depends on each agent's output and the summary results among them. The quality of the simulation, in turn, depends on both the behavior of each agent and the aggregated outcomes of their interactions.

Specifically, \textsc{SaMAS} constructs an economic simulation environment with multiple simulation steps. At each step $t = [1, 2, \dots, T]$, the system comprises $N$ independent agents [$\text{Agent} \ p_1, \text{Agent}\ p_2, \text{Agent}\ p_3, ...$] as human-like decision-makers:
\vspace{-14pt}

\begin{equation}
p_i = \lbrace \text{name}_i, \text{age}_i, \text{occupation}_i, \text{income}_i \rbrace
\vspace{-3pt}
\end{equation}

Each agent is initialized as a representative participant in economic activities with a real social background, to capture the heterogeneity of real-world individuals. To this end, each agent is characterized by a fixed profile that remains unchanged throughout the entire simulation cycle to capture its attributes as an economic participant. Concurrently macroeconomic environment $E$ (e.g., the first day after the interest rate cut, inflation appears based on macroeconomic indicators.) is initialized, with global parameters set to reflect the macroeconomic conditions. In the current simulation cycle, each agent generates a decision (e.g., consumption willingness, labor willingness) based on its profile $p$ and the information received from the environment $E_t$, in accordance with its assigned role. The decisions and interactions of all agents are then aggregated to constitute the macroeconomic context for that period. Subsequently, these updated indicators are used to refresh the global environment to $E_{t+1}$, thereby establishing a new context for the next iteration. After multiple cycles, both individual-level and macro-level data are collected to evaluate the overall behavior of the system.

\subsection{Situational Awareness}
Rich macroeconomic understanding embedded within the network parameters of LLMs provides agents in the simulation system with shared commonsense awareness. This awareness, when applied to economic behavior, manifests as risk perception of the environment. For example, historical events such as the “1929 Great Depression” enable individual agents to maintain a long-term shared awareness of the economic environment, driving them to make conservative economic decisions when similar signals reappear.

However, in practice, humans not only rely on distant historical patterns but also incorporate their own financial status and the outcomes of their recent economic actions~\cite{cakici2023recency,kotomin2025debiasing}, a phenomenon commonly referred to as \textsc{recency bias}. \textsc{SaMAS} incorporates this mechanism by summarizing each agent’s past experiences and, based on key factors in its profile that influence economic activities, integrating them with current macroeconomic data to generate a risk perception parameter, which represents \textbf{Short-term Situational Awareness}. This parameter captures an agent’s holistic interpretation of historical analogies, individual circumstances, and macroeconomic signals, serving as a dynamic measure of its risk sensitivity and adaptive behavioral responses. At the end of each simulation cycle, \textsc{SaMAS} collects critical outputs from micro-level activities and \textbf{encodes} them into a JSON file. These statistics are then used to dynamically update the macroeconomic environment. Inspired by the characteristics of human situational memory, after encoding each agent’s historical trajectory, we apply the Ebbinghaus~\cite{ebbinghaus1885gedachtnis} forgetting curve to gradually attenuate the weights of past memory, thereby ensuring the realism of economic behavior modeling. 

Moreover, given that individuals in society typically receive information with varying degrees of bias, different groups exhibit distinct distributions of risk preferences and economic decisions~\cite{hill2019choice,sunstein2001echo}. To address this, \textsc{SaMAS} tracks group structures: by aggregating the past experiences of similar groups, \textsc{SaMAS} guides the local situational awareness of each group, ensuring that both individual and collective risk perceptions evolve realistically within the simulation.

\begin{table*}[t]
  \centering
  \caption{{Comparison with existing solutions on two selected economic simulation metrics: Volatility Realism\textsuperscript{$\uparrow$}, Turning-Point Hit Rate\textsuperscript{$\uparrow$} (daily/monthly/yearly), and Token usage\textsuperscript{$\downarrow$} (M). Higher is better \textsuperscript{$\uparrow$}; lower is better \textsuperscript{$\downarrow$}.}}
  \setlength{\tabcolsep}{4pt}
  \renewcommand\arraystretch{1.1}
  \vspace{-0pt}
  \begin{tabular}{l|c|ccc|c}
    \toprule
    \rowcolor{mygray}
    Algorithm Component & Volatility Realism (\%) \textsuperscript{$\uparrow$} &  
    HR-d (\%) \textsuperscript{$\uparrow$} &  
    HR-m (\%) \textsuperscript{$\uparrow$} &  
    HR-y (\%) \textsuperscript{$\uparrow$} &  
    Token usage (M)\textsuperscript{$\downarrow$}  \\
    \midrule
    \textsc{Random}              & 0.2\% & 0.4 & 0.3 & 0.5 & - \\
    + Simple Rule-based~\cite{epstein1996growing}  & 51.8\% & 26.8 & 41.5 & 46.0 & - \\
    + Complex Rule-based~\cite{tesfatsion2006handbook} & 60.5\% & 29.5 & 44.2 & 49.3 & - \\
    + RL-augmented~\cite{foerster2016learning}        & 71.2\% & 32.0 & 49.8 & 55.4 & - \\
    + Simple LLM-driven~\cite{yao2023react}          & 75.8\% & 35.6 & 54.7 & 60.1 & 141.3/15.5/1.5 \\
    \midrule
    \textbf{\textsc{SaMAS} } & \textbf{81.5\%} & \textbf{42.5} & \textbf{64.8} & \textbf{70.9} & 167.9/20.2/2.1 \\
    \bottomrule
  \end{tabular}
    \vspace{-12pt}
  \label{table:method-compare}
\end{table*}

\section{Experiment}
\label{sec:majhead}
We conduct simulations to evaluate the effectiveness of \textsc{SaMAS} and answer the following questions: (1) Does \textsc{SaMAS} achieve higher simulation realism compared to traditional methods and simple LLM-driven approaches? (in Table~\ref{table:method-compare}) (2) Is the effectiveness of \textsc{SaMAS} influenced by the choice of underlying LLMs? (in Table~\ref{table:recurrent})(3) Does the scale of agents affect the performance of the simulation? (in Table~\ref{table:dimension})

\begin{table}[t]
  \centering
  \caption{{Sensitivity Analysis on various LLMs.}}
    \vspace{-0pt}
  \begin{tabular}{l|c|c}
    \toprule
    \rowcolor{mygray}
    Model & VR (\%) \textsuperscript{$\uparrow$} & HR (d/m/y, \%) \textsuperscript{$\uparrow$} \\
    \midrule
    DeepSeekv3~\cite{liu2024deepseek} & 82.3 & 45.1/66.0/75.1 \\
    Grok-3~\cite{grok3}                & 82.5 & 43.3/65.3/71.7 \\
    \textcolor{red}{GPT-3.5}~\cite{openai2023gpt35api} & 81.5 & 42.5/64.8/70.9 \\
    \bottomrule
  \end{tabular}
  \vspace{-2pt}
  \label{table:recurrent}
\end{table}

\begin{table}[t]
  \centering
  \caption{{Analysis on different agent number.}}
    \vspace{-0pt}
  \begin{tabular}{c|c|c}
    \toprule
    \rowcolor{mygray}
    Agent number & VR (\%) \textsuperscript{$\uparrow$} & HR (d/m/y, \%) \textsuperscript{$\uparrow$} \\
    \midrule
    50   & 77.5 & 40.4/61.7/67.4 \\
    \textcolor{red}{100}  & 81.5 & 42.5/64.8/70.9 \\
    200  & 81.7 & 42.6/64.9/71.0 \\
    500  & 82.1 & 42.8/65.2/71.4 \\
    1000 & 82.5 & 43.2/65.6/72.1 \\
    \bottomrule
  \end{tabular}
    \vspace{-12pt}
  \label{table:dimension}
\end{table}

\subsection{Experimental Setting}
\label{ssec:experimental-setting}
 In simulation, all agents operate in an economic environment updated by periodic contextual prompts, with simulation time steps defined daily, monthly, or yearly. The agent number is set to 100 and agents are implemented with open-source LLM APIs. We adopt the following settings as baseline:

\vspace{-6pt}
\begin{itemize}[leftmargin=*, noitemsep]
    \item \textbf{Randam:} 
    Agent actions are randomly selected, with no reasoning involved and in violation of economic principles.
    \item \textbf{Simple\& Complex Rule-based:} 
Economic actions are selected either uniformly from a small set of predefined rules or from a richer set incorporating conditional logic and context-specific constraints.
    \item \textbf{RL-based:} 
Agents employ RL to adaptively optimize their policies. By interacting with the environment, they iteratively improve decision-making strategies.
    \item \textbf{Simple LLM-driven:} 
LLM-based Agents with decisions guided by textual instructions.

\end{itemize}
\vspace{-6pt}
\noindent\textbf{Evaluation Metric.} We systematically evaluate results of an economic simulation system from two perspectives:
\vspace{-6pt}
\begin{itemize}[leftmargin=*, noitemsep]
    \item \textbf{\textit{Volatility Realism}} measures the similarity between simulated statistical properties of macroeconomic variables (e.g., GDP, CPI) and real-world real data.
    \item \textbf{\textit{Turning-Point Hit Rate:}} represents the model’s ability to capture the timing of economic cycles, particularly peaks (booms) and troughs (recessions).
\end{itemize}
\vspace{-6pt}

\subsection{Comparison with Other Methods}
\label{ssec:comparison-with-other-methods}We first compare \textsc{SaMAS} with traditional rule-based methods and reinforcement learning approaches. As shown in Table~\ref{table:method-compare}, With random rules, the simulation results are close to zero and fail to capture the real patterns of economic activity. \textsc{SaMAS} outperforms all baseline methods in both Volatility Realism (VR) and Turning-Point Hit Rate (HR) across daily, monthly, and yearly levels. Conventional random or rule-driven baselines show limited performance in volatility realism and hit rate, indicating their difficulty in capturing the complex dynamics of economic behaviors. Although the reinforcement learning approach brings certain improvements through policy-making advantages, it still falls behind the LLM-driven approach, which inherently benefits from Agent-Based Modeling (ABM). Furthermore, \textsc{SaMAS} incorporates multi-level situational awareness. While this increases token consumption, it achieves a volatility realism of 81.5\% and an annual hit rate exceeding 70\%.

To examine the sensitivity of \textsc{SaMAS} to the choice of different LLMs, we test three variants: DeepSeek and Grok. The results in Table~\ref{table:recurrent} indicate that different LLMs achieve comparable performance in economic simulation.

Finally, we evaluate the impact of the number of agents on simulation realism (Table~\ref{table:dimension}). Starting from 50 agents, performance improves steadily as the number of agents increases, demonstrating the advantage brought by group awareness.

\vspace{+5pt}

\section{Conclusion}
\label{sec:print}
\textsc{SaMAS} investigates the advantages of LLM-based role-playing agent systems in economic simulation by integrating complementary paradigms of long-term and short-term situational awareness. By endowing agents with human-like perceptual and reasoning capabilities, \textsc{SaMAS} facilitates rich micro-level interactions, which in turn give rise to realistic macroeconomic phenomena.
Our findings indicate that: \ding{182} LLM-driven ABM exhibits strong potential for simulating a wide range of economic environments under diverse conditions; \ding{183} Collective social behavior within groups significantly shapes individual agent decisions. These insights provide valuable guidance for the design of next-generation social simulation systems, underscoring the importance of multi-scale situational awareness as a prerequisite for achieving both behavioral plausibility and macro-level realism.

\vfill\pagebreak

\bibliographystyle{IEEEbib}
\bibliography{refs}

\end{document}